\documentclass[3p]{elsarticle}

\usepackage{lineno,hyperref}
\usepackage{url,microtype,subcaption}
\usepackage{graphicx}
\usepackage[utf8]{inputenc} %unicode support
\usepackage{amsmath}
\usepackage{amsfonts}
\newcommand{\nlni}{\newline\newline\noindent}
\journal{}

\biboptions{authoryear}

\begin{document}

\begin{frontmatter}

\title{Quantifying changes in the British cattle movement network}

%% or include affiliations in footnotes:
\author[icuhi,sruc]{Andrew J Duncan\corref{mycorrespondingauthor}}
\cortext[mycorrespondingauthor]{Corresponding author}
\ead{andrew.duncan.ic@uhi.ac.uk}

\author[sruc,epic]{Aaron Reeves}

\author[sruc,epic]{George J Gunn}

\author[sruc]{Roger W Humphry}

\address[icuhi]{Inverness College UHI, 1 Inverness Campus, Inverness, IV2 5NA, UK}
\address[sruc]{Epidemiology Research Unit, Department of Veterinary and Animal Science, Northern Faculty, Scotland's Rural College (SRUC), An L\`{o}chran, 10 Inverness Campus, Inverness IV2 5NA, UK}
\address[epic]{Scottish Government's Centre of Expertise on Animal Disease Outbreaks (EPIC)}

\begin{abstract}
  \noindent The Cattle Tracing System database is an online recording system for cattle births, deaths and between--herd movements in the United Kingdom. It is an excellent resource for any researchers interested in networks or modelling infectious disease spread through the UK cattle system. Although it has been thoroughly examined, the most recently reported movement analysis is from 2009. This article uses the database to construct weighted directed monthly movement networks for two distinct periods of time, 2004--2006 and 2015--2017, to quantify by how much the underlying structure of the network has changed. Substantial changes in network structure may influence policy--makers directly or may influence models built upon the network data, and these in turn could impact policy--makers and their assessment of risk. Four general network measures are used (total number of nodes with movements, movements, births and deaths), in conjunction with network metrics to describe each monthly network. Two updates of the database were examined to determine by how much the movement data stored for a particular time period had been cleansed between updates. Statistical models show that there is a statistically significant effect of the time period (2004--2006 vs 2015--2017) in the values of all network measures and six of nine network metrics. Changes in the sizes of both the Giant and Weakly Strongly Connected components predict reductions in the upper and lower bounds of the maximum epidemic size. Examination of the updates of the database show that there are differences in records between updates and therefore evidence of historical data changing between updates. Accurate modelling of disease spread through a network requires representative descriptions of the network. The authors recommend that where possible the most recent available data always be used for network modelling and that methods of network prediction be examined to mitigate for the time required for data to become available.
\end{abstract}


\end{frontmatter}

%\linenumbers

\section*{Introduction}
\noindent
The Cattle Tracing System (CTS) \citep{cts} is an online recording system for cattle births, deaths and between--herd movements in the United Kingdom (UK). It is a legal requirement (since 2001) for cattle owners in the UK to report these details, and failure to report may lead to penalties \citep{bcms-cts, Vernon2011}. Instructions for entering data into the CTS are available online \citep{bcms-cts} and include the information that all movements should be reported to the British Cattle Management Scheme (BCMS) within three days of the movement. The CTS itself was establised in September 1998 \citep{cts-date}. All occasions where an animal moves from one holding and onto another must be recorded as a movement, as must all births and deaths. Movement records such as the CTS data are commonly used to construct static networks where movements recorded over a period of time (week, month or year) are collected together to form a single snapshot. Outside of the UK, similar network analysis of animal movements has been carried out in Brazil \citep{Menezes2020}, Canada \citep{Dube2008}, Germany \citep{Buttner2013, Buttner2014}, Italy \citep{Bajardi2011, Natale2009}, Slovenia \citep{Knific2020}, Sweden \citep{Frossling2012, Lindstrom2009} and Switzerland \citep{Vidondo2018}. Modelling the spread of a disease through networks \citep{Newman2003a} is an additional widespread application of animal movement network data (for example \citep{Craft2011, Eames2002, KeelingEames2005, Martinez2009}). Similar networks are considered in human epidemiology when predicting the behaviour of infectious diseases \citep{Cui2021}.
\nlni
These, and similar, models are underpinned by the network data and, in turn, the models can influence policy-makers or aid in contingency planning for outbreaks \citep{ShirleyRushton2005EpiInf, Kao2007, Dube2009, Vernon2012, Brooks-Pollock2015}. The application of models of disease spread, and more generally of animal movement networks, to inform animal health policy is increasingly common \citep{Mohr2018}. In order for these policies to be credible, the evidence used to inform them likewise must be valid and credible.  This relies, in turn, on up--to--date and accurate data. Herein we present a quantitative assessment of network structure between two distinct periods, to assess whether the overall structure has changed. Substantial changes may influence model results, which in turn could impact policy-makers and their assessment of risk.
\nlni
Within the UK farming industry a good deal of analysis has been conducted on animal movements, particularly following the foot and mouth disease (FMD) outbreaks in 2001 and 2007 \citep{fmd_gov}. The movement data in CTS has previously been described in detail by \citet{Robinson2006} and \citet{Vernon2011}. Both papers examined details including the number of births, deaths and total number of movements. \citet{Robinson2006} examined movements, births and deaths from the CTS database on a daily basis from 2002 until 2005. Their time series analysis showed little overall trend outside of seasonal variation except for a rise in the number of cattle and a small rise in the number of movements off and onto holdings. \citet{Vernon2011} analysed cattle movements from 1999 through to 2009 and showed that again there was no trend in number of movements, either summarised monthly or yearly. Although the years analysed, and the time periods over which the networks are summarised, differ between these two studies, the overall picture is of seasonal variation with small changes over time.
\nlni
In many cases, network metrics \citep{Newman2003a} are used to summarise the network \citep{Dube2009} or identify locations or movements with particular attributes relevant to the spread of disease.  With regard to the movement of animals, the locations that the animals move on and off of become the ``nodes'' of a network, whilst the movements themselves are the ``edges''. The ``betweenness'' metric, the extent that a node lies on paths between other nodes and one of several measures of centrality \citep{Dube2009, Newman2010_Networks}, has been used by others to predict which holdings (\textit{i.e.}, nodes) are important in disease spread  \citep{Christley2005b, , ShirleyRushton2005EcoComp, Natale2009, Rautureau2011, Gates2014a}. \citet{Ortiz2006} used betweenness to identify holdings crucial to the spread of FMD during the 2001 outbreak. It has been shown that eliminating nodes from a network based on betweenness is the fastest way to reduce the size of the ``Giant Strongly Connected Component'' (GSCC), i.e. the collection of nodes which are all connected to one another \citep{Rautureau2011}. This breaks down the entire network into much smaller parts, slowing the spread of disease across the network as a whole.  Eliminating nodes by ``degree'' (the number of movements/edges that a node has) has also been identified as being effective but more restrictive to trade than using betweenness \citep{Natale2009}. \citet{Mweu2013} showed that both metrics are influential as removing Markets (which had the highest median values of both betweenness and degree) had a great impact on the size of the GSCC in Denmark. Whilst, in the French cattle network, isolating the nodes with the top 5\% of betweenness values, broke up the GSCC \citep{Dutta2014} and in the Slovenian cattle network both degree and betweenness could be used to reduce the GSCC but total degree was more effective \citep{Knific2020}.
\nlni
Movement networks (and the data used to construct them) have been used to inform or analyse policy for some time \citep{ShirleyRushton2005EpiInf, Kao2007, Vernon2012} and the timeliness of data availability and quality has already been highlighted \citep{Dube2009, Brooks-Pollock2015}. Futhermore it has been advised that epidemics be modelled in advance of an outbreak \citep{Taylor2003, ShirleyRushton2005EpiInf} and whilst the CTS data is an excellent resource, it takes time for the databases available to be updated. This means that modelling often takes place with data that is a year or more old. If the underlying network structure has changed significantly from historical data then this could present problems for any model built on that data. The most recently reported movement analysis is from 2009 \citep{Vernon2011} and more up to date records are now available for examination.
\nlni
The purpose of this article is to use CTS data to construct monthly movement networks for two distinct periods of time, 2004 -- 2006 and 2015 -- 2017, to establish if the structure of the cattle network in the UK has changed. A variety of general measures are used, in conjunction with network metrics to describe each monthly network. In addition to the differences in movement networks between time periods, the quantity of several types of unexpected movements (movements leaving slaughterhouses and movements leaving then returning to a holding with no other holding involved) were examined and two updates of the CTS database were used to see whether the data for the 2004 -- 2006 time period has been amended between updates. Any changes, depending on what they are, could inform us as to how important it is for epidemiologists and modellers to have as up to date movement data as possible. Overall uncertainty presented to policy makers could occur from natural stochasticity which is modelled within the system, or from uncertainties in the structure of the model. These uncertainties could be due to using inaccurate data for constructing the cattle movement network. Reducing this second form of uncertainty (i.e. that associated with the uncertainty in the accuracy of the network data), will lead to reduced overall uncertainty which, in turn, could help risk managers by reducing the range of likely scenarios to consider.

\section*{Materials and Methods}
\subsection*{Construction of Networks}
\noindent
Monthly networks were constructed using records from the CTS database \citep{Mweu2013, Vidondo2018}. For each movement the following information was extracted:
\begin{itemize}
 \item ID number of the animal;
 \item Date of the record;
 \item Whether the record indicates the birth, death, or movement of an animal;
 \item Holding information (county parish holding (CPH) number, location id, location description, holding description and premise description) for both the off (leaving) and on (arriving) holdings, if applicable.
\end{itemize}
\noindent
The CTS database classfies the individual holdings into groups according to three descriptive variables:
\begin{enumerate}
 \item Location description mainly allows us to distinguish between Agricultural Holdings or Slaughterhouses.
 \item Holding description has five categories (Beef, Dairy, Mixed, Not Matched and Other), as determined by the UK Animal and Plant Health Agency (APHA) based on on-farm cattle breed composition (APHA, personal communication).
 \item Premises description is a much finer classification system including Agricultural Holdings, Landless Keepers, Markets, Showgrounds and Slaughterhouses (Red and White Meat) amongst other smaller groups.
\end{enumerate}
\noindent
Having extracted the movement data, the following method was used to construct our monthly networks:
\begin{enumerate}
 \item All records for each calendar month (excluding births, deaths or involving holdings with invalid CPHs/missing location ids) were extracted from the CTS database;
 \item A contact network was constructed for each month \citep{Newman2003a, KeelingEames2005}. The individual holdings became the nodes and the animal movements an edge, meaning the networks were weighted, with the weight of each edge being the number of animals moved between holdings per month. As the movements could be either in or out of the holding, a directed network was used. The weighted, directed networks were constructed using the igraph package \citep{igraph} in the statistical software R \citep{RCore};
 \item Every weighted directed monthly network was summarised through the calculation of general network measures and network metrics \citep{Newman2003a};
 \item The monthly summaries were gathered into two distinct periods: 2004 -- 2006 and 2015 -- 2017 for analysis.
\end{enumerate}
\noindent
The two three-year periods 2004 -- 2006 and 2015 -- 2017 were chosen for several reasons. A three year period was chosen as it was felt by the authors that a single year may not present an accurate sample of the cattle movements, likewise just a two year period. used a three year period up to and including the last full calendar year before the 2007 FMD outbreak \citep{fmd_gov}. A three year period is the minimum needed to observe any exceptional year. As can be seen in figure \ref{fig:total_nodes} in which 2007 is included for reference, this particular year contains some abnormal months associated with the FMD outbreak, which would have adversely influenced the comparison. To match the early three-year period, the most recent complete three-year period available at the time of writing was used. A longer period was not used because the aim was to compare two separate periods rather than observe any particular trend within one of the time periods.
\nlni
Only actual movements of animals were treated as edges in the network: although births and on-farm deaths are recorded in the CTS database, these do not represent ``connections'' between holdings and cannot spread disease directly, thus they were omitted from characterisation of the network. We also omitted all movements where the CPH of either holding was invalid (contained a parish code of 999) or where the location id was missing. All locations without a location id also lacked a CPH and so could not be identified. This would render contact tracing impossible \citep{Kiss2006a}. A plot of the number of omitted movements due to invalid CPHs and missing location ids can be found in figure S1 in the supplementary information with the number of included movements for comparison. Unlike previous studies \citep{Woolhouse2005}, we wished to include all short-stay locations \citep{Robinson2007b} (\textit{e.g.}, Markets, Showgrounds, \textit{etc.}) as nodes in the network: movements to and from such locations were included in the network characterisation.
\nlni
For each monthly network, within a given period 2004 -- 2006 and 2015 -- 2017, all of the holdings with any off (\textit{i.e.}, outgoing) movements in that three year period were included in the network -- as these nodes could spread disease via animal movements. This ensured that the nodes remained static but did mean that, in any given month, networks may have included nodes with zero movements in that month. To take account of the possibility that any holdings had ceased trading or new holdings had been introduced, the nodes included in the networks were allowed to change between 2006 and 2015. Originally, it was planned to remove all Slaughterhouses as it was assumed that they represented disease cul--de--sacs. However, when processing the data it was noticed that some movements were recorded as leaving holdings classed as Slaughterhouses. These were retained in the networks and movements leaving Slaughterhouses analysed separately in addition to their inclusion.
\newline
\subsection*{Network Analysis}
\noindent
The weighted directed contact networks were summarised using network measures, general movement based calculations, and network metrics \citep{Newman2003a}. The network metrics can be further divided into two types, network level and node level. The network level metrics, and the general network measures, produce a single value for the entire monthly network whilst the node level metrics provide values per node per month. The network measures calculated were the number of births and deaths per month (not just those from nodes present in our networks) along with total number of movements and total number of nodes with movements in each of our monthly networks.
\newline
\subsubsection*{Network Level Metrics}
\noindent
The network level metrics shown below all produce a single value for each monthly network.
\begin{itemize}
 \item Assortativity. Assortativity is a measure of the tendency of nodes within a network to have connections with similar (or dissimilar) nodes \citep{Newman2003b}. Similarity in this case is based on degree value. In our case a positive value of assortativity means that holdings with a high number of movements would be sending/receiving animals to/from other holdings with a high number of movements. A negative value of assortativity would mean the opposite, holdings with a high number of movements would be sending/receiving animals to/from other holdings with a low number of movements. Degree assortativity can be used as an indicator of the susceptibility of a network to random or targeted node isolation procedures, effectively removing them from the network. A dissassortative network (negative assortativity value) is much more vulnerable to targeted node isolation than an assortative network \citep{Newman2003b}. It is expected that disease spread will be quicker on an assortative network than a dissassortative one \citep{Kao2007}.
 \item Average Path Length. The average path length is the shortest path (fewest number of movements) between two holdings, averaged over the entire monthly network \citep{Dube2009}. It is calculated across the entire monthly network and any disconnected nodes will not contribute. A pair of nodes, not connected to the rest of the network, that are themselves connected ($A \rightarrow B$, a disconnected dyad) will contribute a path length of 1 \citep{igraph}. Shorter path lengths suggest that disease spread will be quicker. In turn this may require stricter containment methods \citep{KeelingEames2005}.
 \item Sizes of the Giant Strongly Connected Component (GSCC) and Giant Weakly Connected Component (GWCC). In a directed network the GSCC is the largest group of nodes that are mutually connected. That is every node has a connection to and from every other node. The GWCC is the GSCC plus all nodes that have edges into and out from it. In our case the GWCC is the GSCC plus all holdings that send animals into or receive animals from it. The size of all strongly and weakly connected components on the monthly networks were found and the maximum sizes of both component types used to describe the network. Assuming there are no changes or interventions, the size of the GSCC represents a reasonable lower bound on the maximum size of a final epidemic whilst the size of the GWCC provides an upper bound \citep{Kao2009} which is demonstrated empirically in figure \ref{fig:gscc_gwcc_epi}. In both networks shown in figure \ref{fig:gscc_gwcc_epi} the maximum size of giant strongly connected components (GSCC) is three with the triangle of nodes all connected to one another. Similarly the size of the giant weakly connected component (GWCC) is four as they both have an additional node connected into or out of the GSCC. If we do not know in advance the direction of the connection of the additional node, we do not know which case we have. In the top network, the additional node is connected into the network and cannot be infected if the epidemic starts on the GSCC. In this case the maximum epidemic size is the size of the GSCC. In the lower network, the additional node is connected out of the network and is part of an epidemic starting on the GSCC. Here the maximum epidemic size is the size of the GWCC. Targeting the nodes with highest degree value in a dissassortative network quickly reduces the Giant Strongly Connected Component (GSCC) \citep{Mweu2013}, thereby reducing the lower bound of the maximum epidemic size. In our description of the cattle movement network, we present the size of the GSCC and GWCC as a proportion of the nodes present in the monthly networks for each time period.

 \begin{figure}[!htbp]
  \centering{
   \includegraphics[angle = -90, width = 0.8\textwidth]{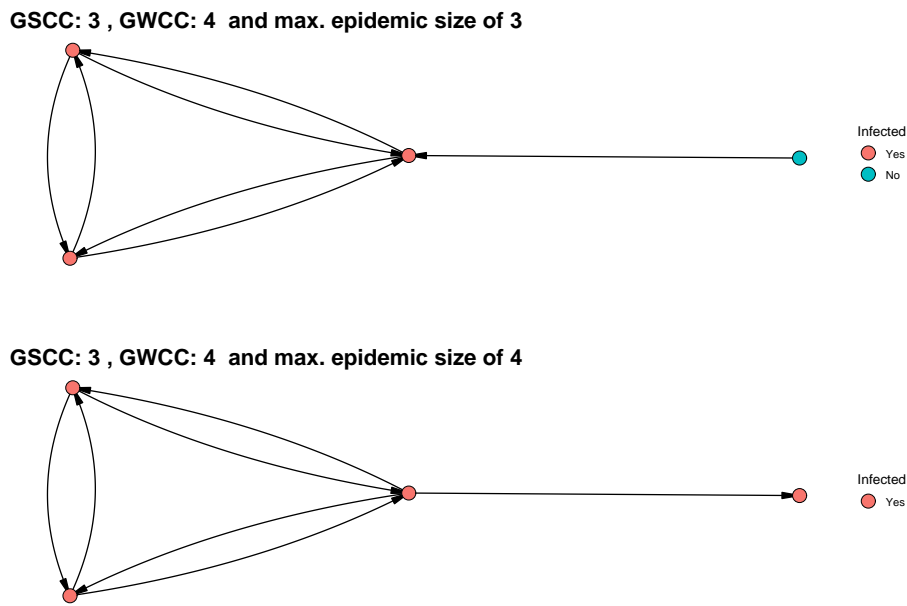}
  }
  \caption{Plot of two networks with identical GSCC and GWCC sizes but different epidemic sizes to exmplify how the sizes of the GSCC and GWCC relate to bounds on epidemic size. Red nodes indicate infection whilst blue nodes are not infected. Both of these networks have had all edge weights set to 1.}
  \label{fig:gscc_gwcc_epi}
 \end{figure}

 \item Proportion of nodes with zero betweenness. Having performed our calculations it was clear that the values of betweenness (see below) had a multimodal distribution with a large number of holdings (nodes) with a betweenness of zero. Distributions such as this are not well summarised by a single statistic (e.g. mean) and are much better described by separating the modes (or ``components''). The holdings with a betweenness value of zero were separated from the others and the proportion of these nodes in the networks was used as an additional network level metric. Distributions and the values of skewness for each monthly network can be seen in the supplementary information.
 \item Reciprocity. The reciprocity of a directed network is the proportion of edges $A \rightarrow B$ for which the reciprocal edge $B \rightarrow A$ also exists \citep{Newman2010_Networks}. A low value of reciprocity shows a highly directional network, meaning that trying to approximate with an undirected network would not be appropriate.
\end{itemize}
\medskip
\subsubsection*{Node Level Network Metrics}
\begin{figure}
 \includegraphics[angle = -90, width = 0.95\textwidth]{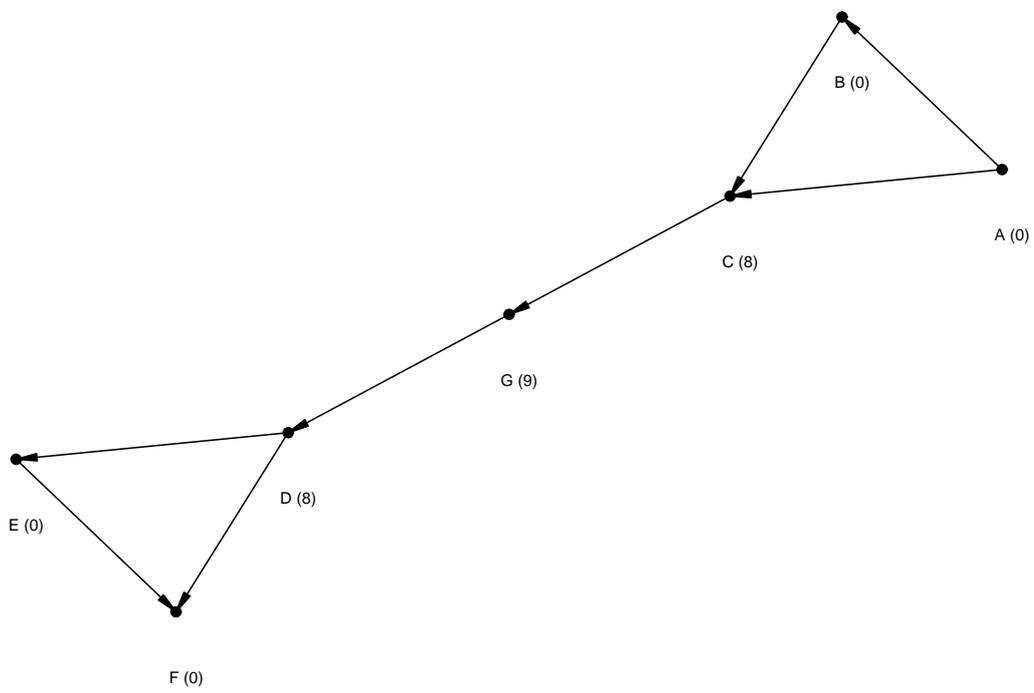}
 \caption{An example network with edge weight of one. The betweenness of each node is shown in brackets.}
 \label{fig:betweenness_network}
\end{figure}

\noindent
Three node level metrics were calculated, betweenness, degree and strength. The betweenness of a holding is the number of times it lies on the shortest path between two other holdings \citep{Dube2009}. Nodes with high betweenness can be thought of as bridges between groups of nodes, connecting nodes that would otherwise not be connected. An example network is shown in figure \ref{fig:betweenness_network} with the betweenness of each node displayed in brackets. Node $G$ could be thought of as a bridge between the two groups and so has the highest value of betweenness, all shortest paths from $A$, $B$ and $C$ to $D$, $E$ and $F$ run through it. Figure \ref{fig:betweenness_network} might also exemplify why it has been shown that targeting (i.e. preventing animal movements on and off) nodes with high betweenness is an efficient way to reduce epidemic size \citep{Rautureau2011, Mweu2013, Dutta2014}, knocking out any of nodes $G$, $C$ or $D$ would break the network into two distinct groups preventing infection, via animal movement, from one group to another. Generally, high betweenness is thought to be a property of nodes with high contribution to the spread of an epidemic \citep{Ortiz2006}.

\noindent
\nlni
In our weighted networks the degree of a holding (node) is the number of other holdings it had to connections to per month. As direction of the connection is accounted for, the animals can move onto or off the holdings, three different degree values were calculated per holding, per monthly network. For each holding the ``degree in'' represents the number of other holdings it received animals from, the ``degree out'' is the number of other holdings it sent animals too and ``total degree'' represents the sum of ``degree in'' and ``degree out'' \citep{Newman2010_Networks}.
\nlni
Whilst degree shows the number of holdings connected, ``strength'' details the number of animals moved between holdings. The strength of a holding (node) is sum of the weights of its edges \citep{Barrat2004}. In our network the edge weight represented the number of animals and thus the strength denotes the total number of animals moved. As with degree, three different strength values were calculated per holding, per monthly network -- depending on the direction of the animal movement. The ``strength in'' represents the sum of all edge weights onto a node, the total number of animals moving onto the holding and the ``strength out'' is the sum of all edge weights off of a node (i.e. the total number of animals moving off of a holding). The ``total strength'' is the sum of ``strength in'' and ``strength out''.
\subsubsection*{Effect of Period}
\noindent
Along with overall changes in the values of the network measures and metrics we were interested in establishing whether period (either 2004 -- 2006 or 2015 -- 2017) had a significant effect on the values. For both the general network measures and the network metrics, mixed effect models (\ref{eq:lm}) were used to assess whether period had a statistically significant effect. A model was fitted to each network measure and metric. In all models the fixed effects were the dependent variable (network measures or metrics) lagged by one month $(y_{t-1})$; period (values of `One' or `Two') and month (categorical values of January to December as the monthly effect was not linear). $\gamma_k$ is the random effect of the $k^{th}$ year with $\gamma_k \sim N(0, \sigma_{year}^{2})$ and $\epsilon_t \sim N(0, 1)$.
\begin{align}
 y_{t} \sim \beta _0 + \beta_1 y_{t-1} + \beta_2\textrm{month}_t + \beta_3\textrm{period}_t + \gamma_{k} + \epsilon_{t} \quad & \{t \in \mathbb{N}:\, 2 \leq t \leq 36\}\label{eq:lm} \\
 & \{k \in \mathbb{N}:\, 1 \leq k \leq 6\} \nonumber
\end{align}
\noindent
For the network level metrics and network measures, we had a single value for each monthly network and these were used directly. However, the node level metrics of (non--zero) betweenness, degree and strength had a value per node, per month. In order to model their values these were reduced to a single value per monthly network. Upon investigation, the distributions of all metrics were skewed with a tail to the right and so the models used the median value per month. The skewness values for the distributions of the three node level metrics can be seen in figures S2 -- S4 in the supplementary information.
\newline
\subsection*{Problems with the Network}
\noindent
When initially constructing the networks and doing some data preparation, we noticed two types of movement that were not expected to appear (hereafter referred to as ``problem'' movements). The first was mentioned above: movements off (animals leaving) Slaughterhouses. The other that we examined were movements which leave and return to the same location but are recorded without a short stay location (for example a market, where the animal would be for a short period of time). These edges are sometimes referred to as self--edges or loop edges. We examined how these differed between our two time periods but also how they changed between CTS updates.
\newline
\smallskip
\subsection*{Differences between CTS updates}
\noindent
Updated data is provided by the Animal and Plant Health Agency (APHA) Rapid Analysis and Detection of Animal-related Risks (RADAR) team to the Scottish Government's Centre of Expertise on Animal Disease Outbreaks (EPIC) on an approximately quarterly basis.  These quarterly updates include newly recorded movements, births, and deaths from the last three months.  Additionally, on an annual basis, APHA RADAR provides a complete updated version of the CTS database, which also incorporates corrections to historical data.  We examined two such annual updates to determine the extent of the corrections made and to identify the nature of these changes. The first contained all movements from 1$^\textrm{st}$ January 2001 until 14$^\textrm{th}$ February 2014 and was only used for the comparison of updates. The second update, which was used for all analysis including the comparison between updates, contained all movements from 1$^\textrm{st}$ January 2001 up to 31$^\textrm{st}$ December 2017. The differences in the types of ``problem'' movement were examined between these two updates along with checking how many alterations there had been to the way the holdings (nodes) are described -- CPH, location description, holding description and premise description.
\nlni
All analyses were carried out using the statistical software R \citep{RCore} with the network metrics calculated using the igraph package \citep{igraph} and plotted using the ggplot2 package \citep{ggplot2}. The mixed effect models were calculated using the lme4 and lmerTest packages \citep{lme4, lmerTest}.

\section*{Results}
\subsection*{Network Analysis}
\subsubsection*{General Network Measures}
\begin{figure}
 \includegraphics[angle = -90, width = 0.95\textwidth]{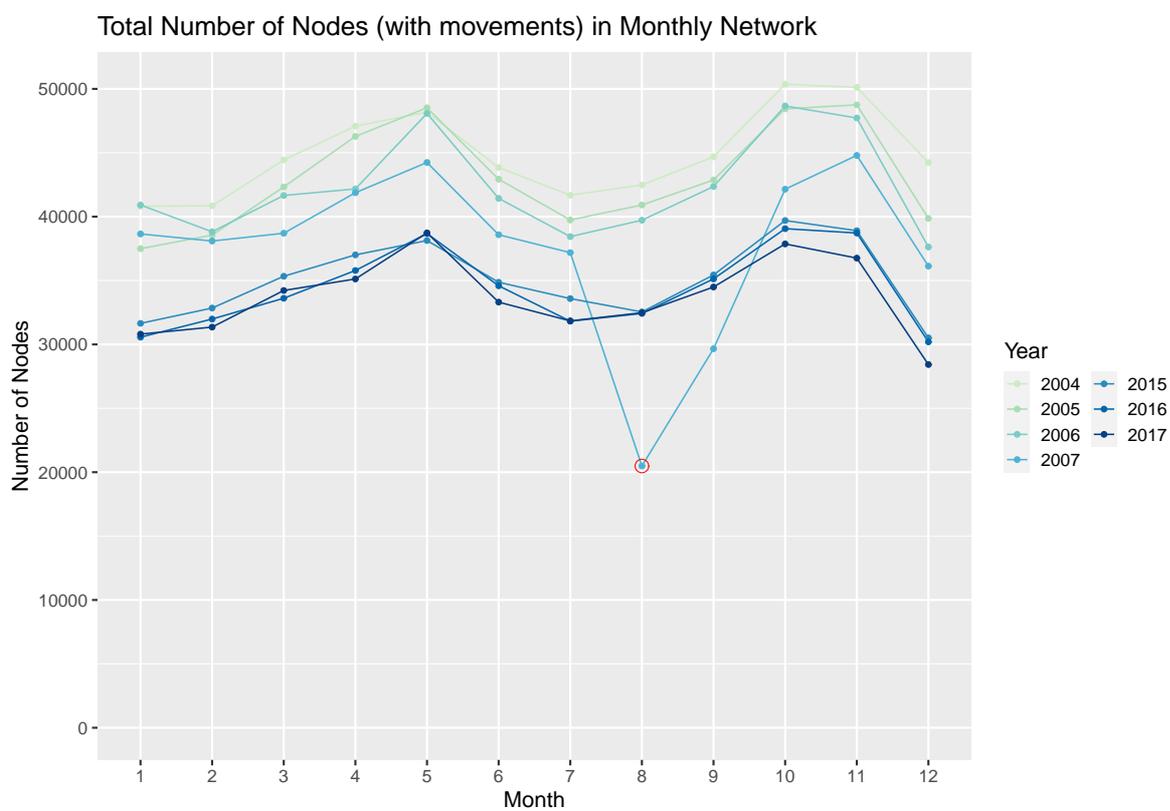}
 \caption{Plot of the number of nodes (with movements) in the monthly networks. The lines and points become darker as the networks become more recent. The decrease in August 2007 associated with the FMD outbreak is highlighted.}
 \label{fig:total_nodes}
\end{figure}

\begin{figure}
 \includegraphics[angle = -90, width = 0.95\textwidth]{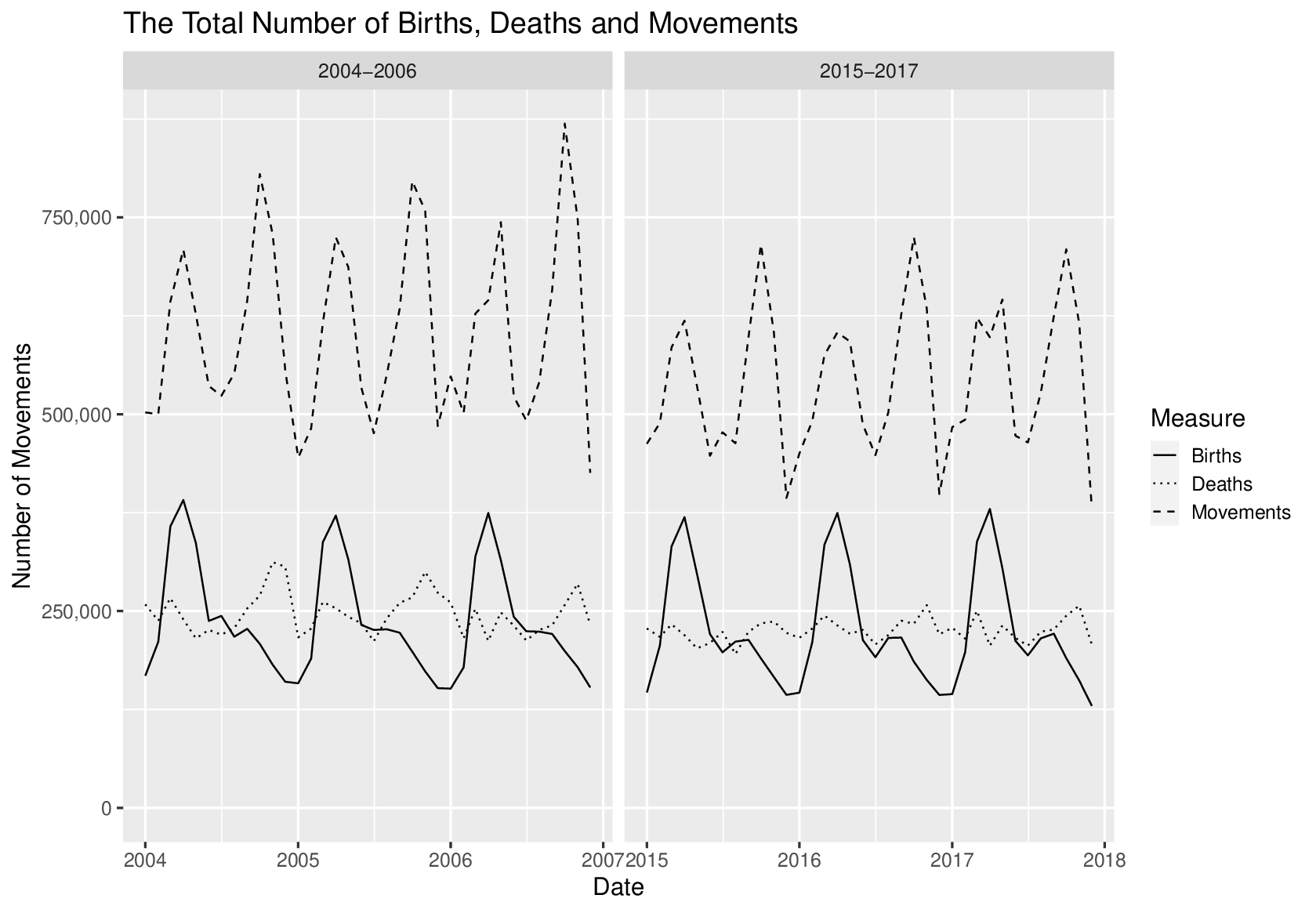}
 \caption{Plot of the total number of Births (solid), Deaths (dotted) and Movements (dashed) per month for each time period.}
 \label{fig:move_birth_death}
\end{figure}

\noindent
The first step we completed in examining the network and the difference between the two time periods was to look at some general attributes of the network. Figure \ref{fig:total_nodes} shows the total number of nodes that had movements, per month, for all the years we examined, whilst figure \ref{fig:move_birth_death} shows the total number of births, deaths and movements in the network for both time periods. The number of deaths includes deaths at Slaughterhouses that are not considered in our normal monthly movement network. Both figures show seasonality within the year. The networks for 2007, including the last FMD outbreak, are included in figure \ref{fig:total_nodes} for comparison and it is clear how the months of August (highlighted) and September differ to those in other years.

\noindent
Table \ref{tab:num_nodes_per_period} shows the number of nodes with movements (those nodes present in the networks) whilst table \ref{tab:totals_per_year} shows the number of movements in the network along with the numbers of births and deaths within the CTS database for each year within the two time periods. Both tables show decreases between the two periods. Monthly values for the general network measures can be seen in tables S1, S2 and S3 of the supplementary information.
\begin{table}[ht]
\centering
\begin{tabular}{cc}
  \hline
2004 -- 2006 & 2015 -- 2017 \\
  \hline
89,521 & 75,625 \\
   \hline
\end{tabular}
\caption{Total number of nodes with at least one off movement
             for each three year period.}
\label{tab:num_nodes_per_period}
\end{table}

\begin{table}[ht]
\centering
\begin{tabular}{r|lll|lll}
  \hline
Measure & 2004 & 2005 & 2006 & 2015 & 2016 & 2017 \\
  \hline
Movements & 7.33 & 7.19 & 7.32 & 6.39 & 6.53 & 6.64 \\
  Births & 2.94 & 2.80 & 2.78 & 2.69 & 2.70 & 2.69 \\
  Deaths & 3.03 & 2.99 & 2.87 & 2.64 & 2.74 & 2.71 \\
   \hline
\end{tabular}
\caption{Total number (millions) of movements in the network
             along with total births and deaths in the CTS database for each
             year in both time periods.
             Unrounded values for each month are available in tables 1, 2 and 3
             of the supplementary information.}
\label{tab:totals_per_year}
\end{table}

\subsubsection*{Network Level Metrics}
\begin{figure}
  \includegraphics[angle = -90, width = 0.95\textwidth]{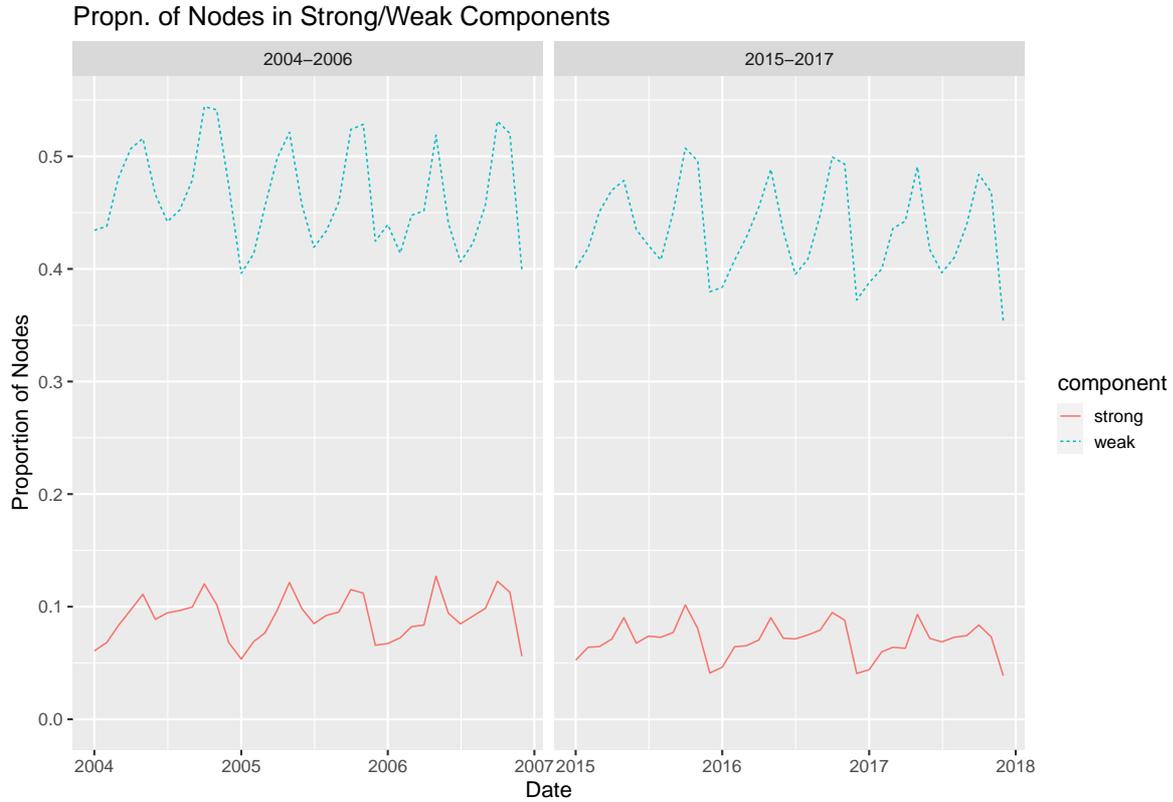}
 \caption{Plot of the proportion of nodes in the network present in the maximum sizes of the giant strongly (red, solid) and weakly (blue, dashed) connected components of each monthly network.}
 \label{fig:strong_weak}
\end{figure}

\noindent Figure \ref{fig:strong_weak} shows the proportion of nodes present in the network that were contained within the GSCC and GWCC for each monthly network. There is a general decrease in the proportion of nodes in both components between the two time periods, along with seasonality. Plots for assortativity, average path length, the absolute sizes of the GSCC and GWCC, proportion of nodes with zero betweenness and reciprocity can be seen in the supplementary information. Similarly to the decreases in proportion of nodes in the GSCC and GWCC, there is an observable decrease in assortativity. However, much of the observable decreases in the absolute GSCC and GWCC is not mirrored in figure \ref{fig:strong_weak}, showing that much of it is due to the reduction in number of nodes between time periods.

\subsubsection*{Node Level Metrics}
\begin{figure}
 \includegraphics[angle = -90, width = 0.95\textwidth]{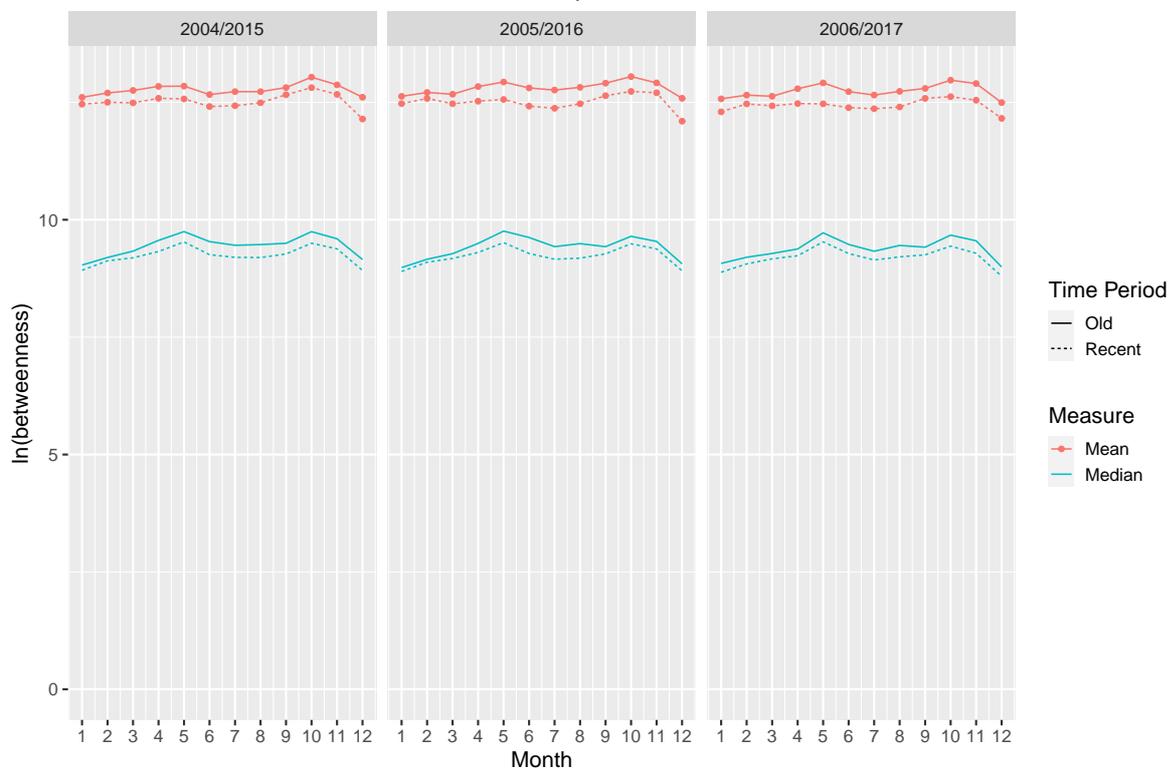}
 \caption{Plot of the mean (red with points) and median (blue) betweenness values for the monthly networks of both the older (solid lines) and more recent (dashed lines) time periods.}
 \label{fig:betweenness_med_av}
\end{figure}

\begin{figure}
 \includegraphics[angle = -90, width = 0.95\textwidth]{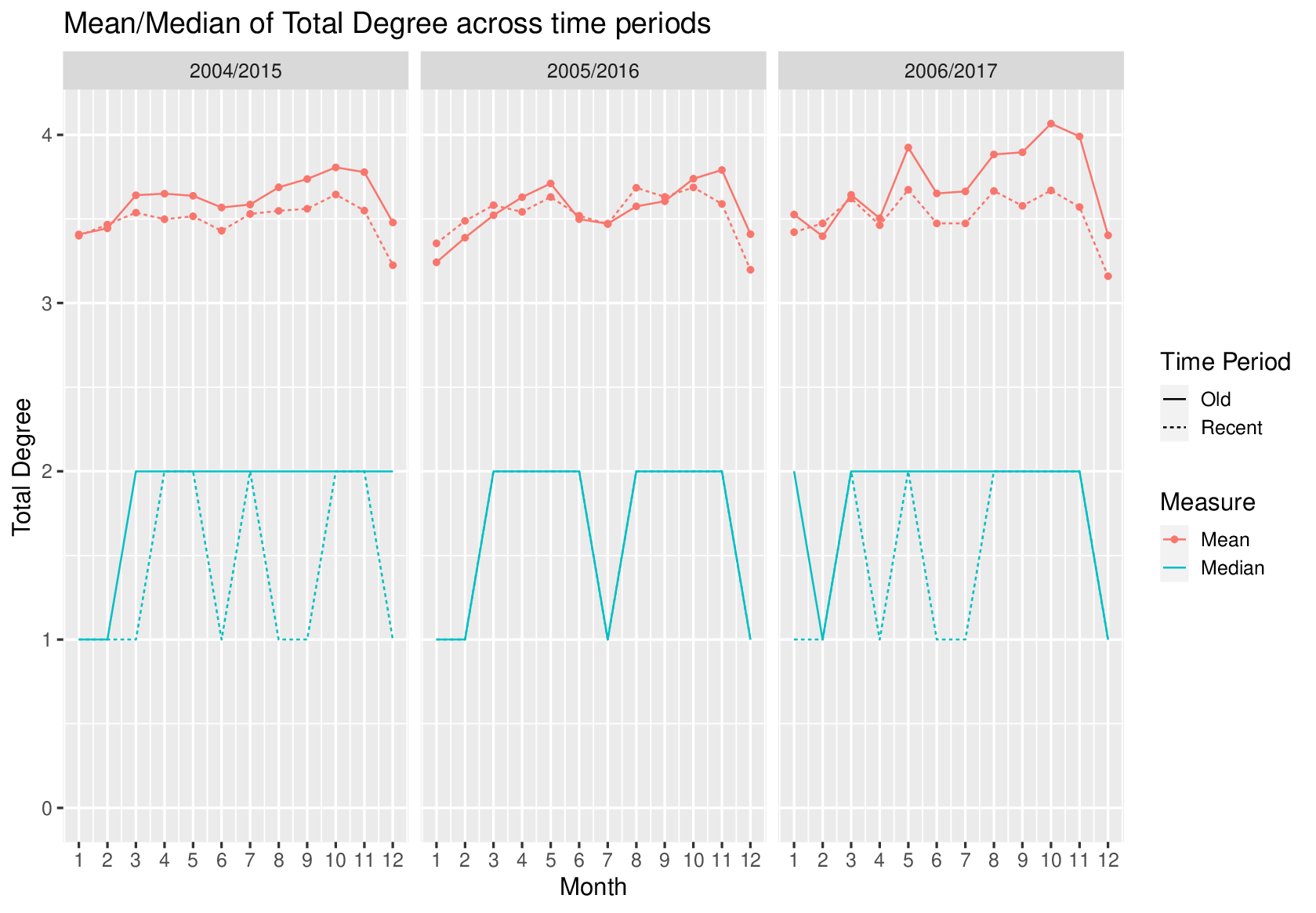}
 \caption{Plot of the mean (red with points) and median (blue) total degree values for the monthly networks of both the older (solid lines) and more recent (dashed lines) time periods.}
 \label{fig:degree_med_av}
\end{figure}

\begin{figure}
 \includegraphics[angle = -90, width = 0.95\textwidth]{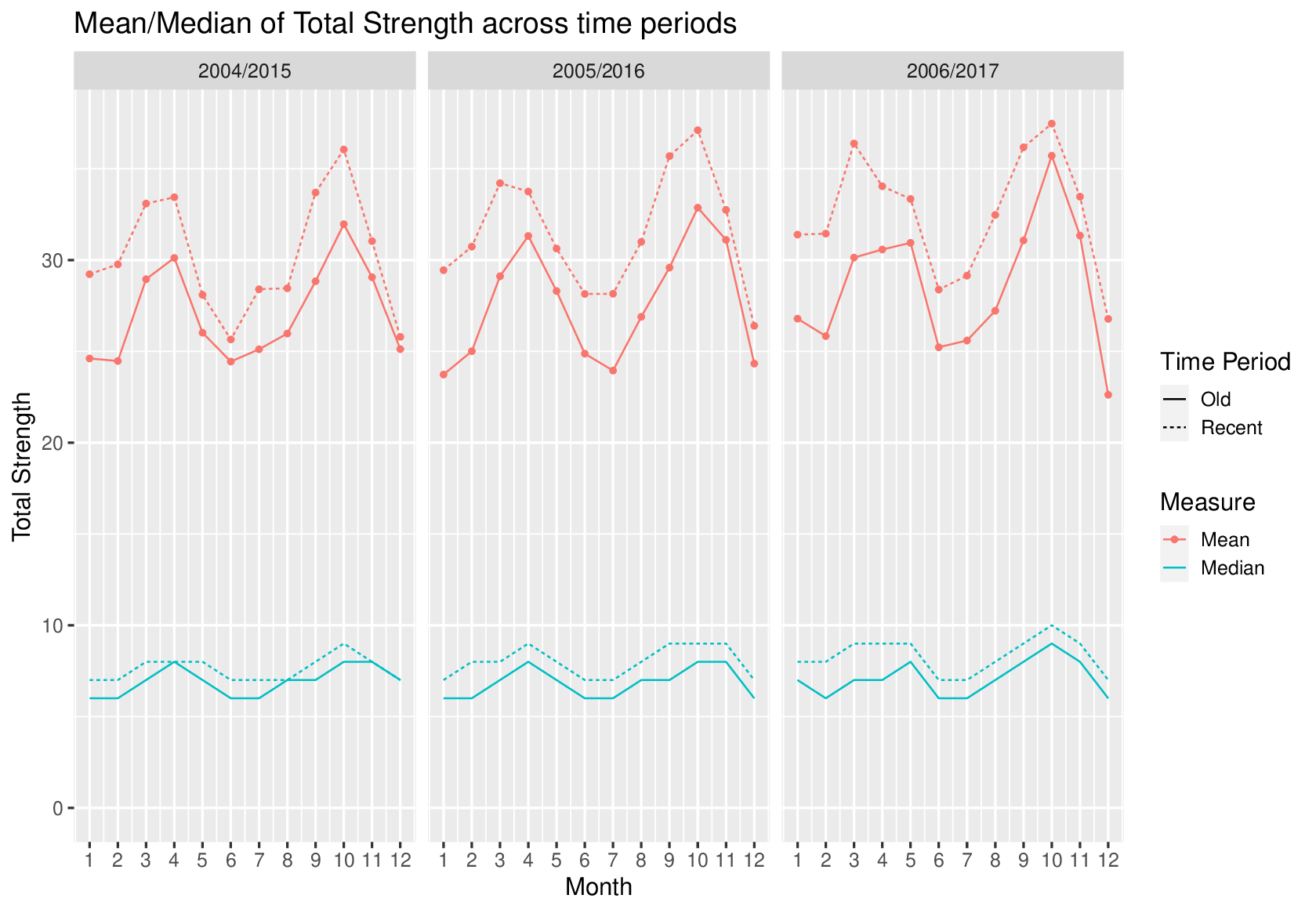}
 \caption{Plot of the mean (red with points) and median (blue) total strength values for the monthly networks of both the older (solid lines) and more recent (dashed lines) time periods.}
 \label{fig:strength_med_av}
\end{figure}

\noindent Figure \ref{fig:betweenness_med_av} shows the median and mean betweenness values of the monthly networks (after all zero valued nodes were removed). They are plotted on a log scale as the distributions are highly skewed with a tail to the right. As with betweenness, degree and strength both produce a value for each node, in each monthly network. Figures \ref{fig:degree_med_av} and \ref{fig:strength_med_av} show the median and mean values of total degree and total strength respectively for each monthly network. These distributions are skewed, as were the betweenness values. Figures \ref{fig:betweenness_med_av} and \ref{fig:degree_med_av} shows that in general there is a decrease between mean/median values of both betweenness and degree in the monthly networks between our two time periods. Contrastingly, figure \ref{fig:strength_med_av} shows a general increase in mean/median strength value probably reflecting an increased number of movements per holding which in turn may be a consequence of a tendency towards greater trading or an increase in holding size or a combination of both.

\subsubsection*{Effect of Period}
\noindent
The effect of period on the values of the general network measures, network level metrics and node level metrics was analysed using mixed effect models (\ref{eq:lm}) and the results can be seen in table \ref{tab:lm_results}. The estimate for the effect of period is included, along with a 95\% confidence interval and p--value, with the intercept for scale.
\begin{table}[ht]
\centering
\begin{tabular}{p{4cm}lllp{2cm}}

 & Estimate & 95\% C.I. on estimate & p-value & Intercept (for scale) \\
  \hline
  Network Measure & & & & \\
  \cline{1-1}
  Total Nodes & -7616 & (-10391, -4841.3) & $1.42 \times 10^{-5}$ & 33475 \\
  Movements & -61016 & (-84627, -37405) & $4.75 \times 10^{-6}$ & 491263 \\
  Births & -7313 & (-12914, -1711.4) & $6.37 \times 10^{-2}$ & 95080 \\
  Deaths & -18219 & (-28531, -7907) & $9.38 \times 10^{-3}$ & 198563 \\
  \hline
  Network Level Metric & & & & \\
  \cline{1-1}
  Assortativity & -0.00492 & (-0.02, 0.009) & $5.36 \times 10^{-1}$ & -0.2544 \\
  Average Path Length & 0.0833 & (-0.02, 0.2) & $3.48 \times 10^{-1}$ & 2.739 \\
  Strong Component & -0.0165 & (-0.02, -0.01) & $7.67 \times 10^{-7}$ & 0.04943 \\
  Weak Component & -0.0279 & (-0.05, -0.009) & $3.44 \times 10^{-2}$ & 0.3831 \\
  Propn. Nodes with Zero Betweenness & 0.0163 & (0.009, 0.02) & $2.37 \times 10^{-3}$ & 0.7081 \\
  Reciprocity & 0.000798 & (-0.003, 0.005) & $7.07 \times 10^{-1}$ & 0.06664 \\
  \hline
  Node Level Metric & & & & \\
  \cline{1-1}
  Median Betweenness & -1080 & (-1670.6, -487.99) & $2.03 \times 10^{-3}$ & 3918 \\
  Median Total Degree & -0.32 & (-0.6, -0.09) & $4.75 \times 10^{-2}$ & 1.692 \\
  Median Total Strength & 1.09 & (0.5, 1.6372) & $8.87 \times 10^{-3}$ & 6.822 \\
\end{tabular}
\caption{Table of the estimate on the effect of period, 95\% CI on the estimate and the
             p-value for the effect of period (either 2004--2006 or 2015--2017)
             in the mixed effect models (defined by \ref{eq:lm}) for each network measure, network level metric and median value of the node level metrics. Year was included as a random effect. The intercept is
             provided for scale.}
\label{tab:lm_results}
\end{table}
\noindent
For the network measures of i) total number of nodes with movements; ii) movements; iii) deaths, there was a clear period effect. However for births, period did not represent a significant effect. In each case the values from 2015 -- 2017 are lower than those from 2004 -- 2006. For all network metrics (network level and node level), except average path length, assortativity and reciprocity there was a period effect.
\nlni
The values of GSCC, GWCC, median betweenness and median total degree all show lower values in 2015 -- 2017 than 2004 -- 2006. In comparison, the values of median total strength and the proportion of nodes with zero betweenness show higher values in 2015 -- 2017. A quantification of the differences can also be provided by the proportional differences between time periods (effect size relative to intercept). For median betweenness, median total degree and median total strength, these proportional differences between time periods were $-0.276$, $-0.189$ and $0.160$ respectively.
\nlni
\subsection*{Problems with the Network}
\begin{figure}
 \includegraphics[angle = -90, width = 0.95\textwidth]{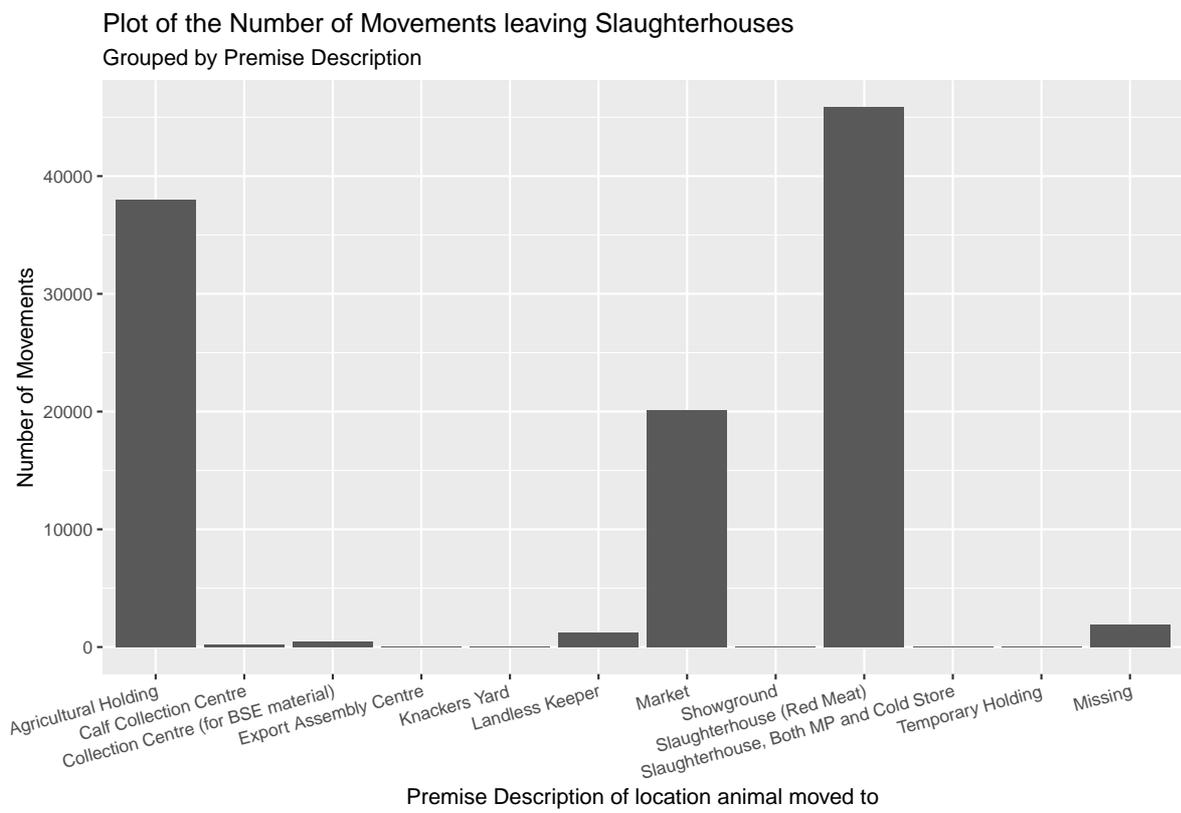}
 \caption{Plot of the Number of Movements leaving Slaughterhouses split by premise type}
 \label{fig:num_off_slaughter}
\end{figure}

\begin{figure}
 \includegraphics[angle = -90, width = 0.95\textwidth]{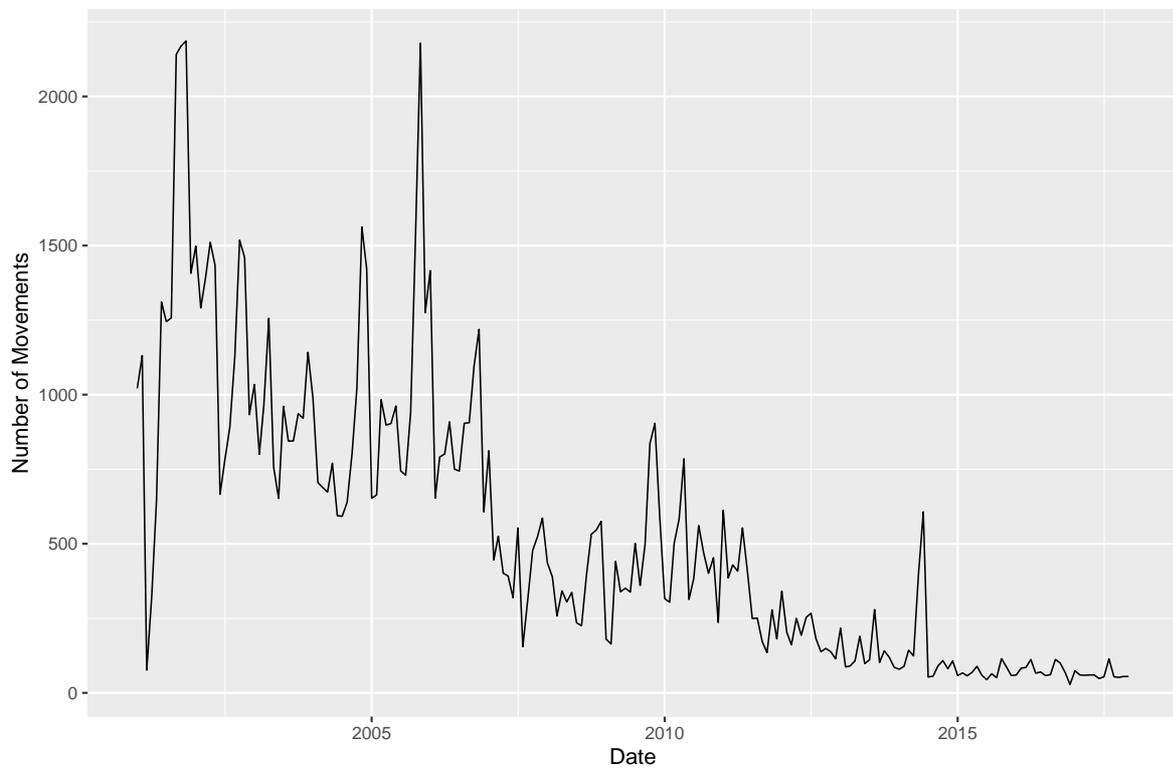}
 \caption{Count of number of movements off Slaughterhouses per month throughout the entire CTS database.}
 \label{fig:off_slaughter_overtime}
\end{figure}

\begin{figure}
 \includegraphics[angle = -90, width = 0.95\textwidth]{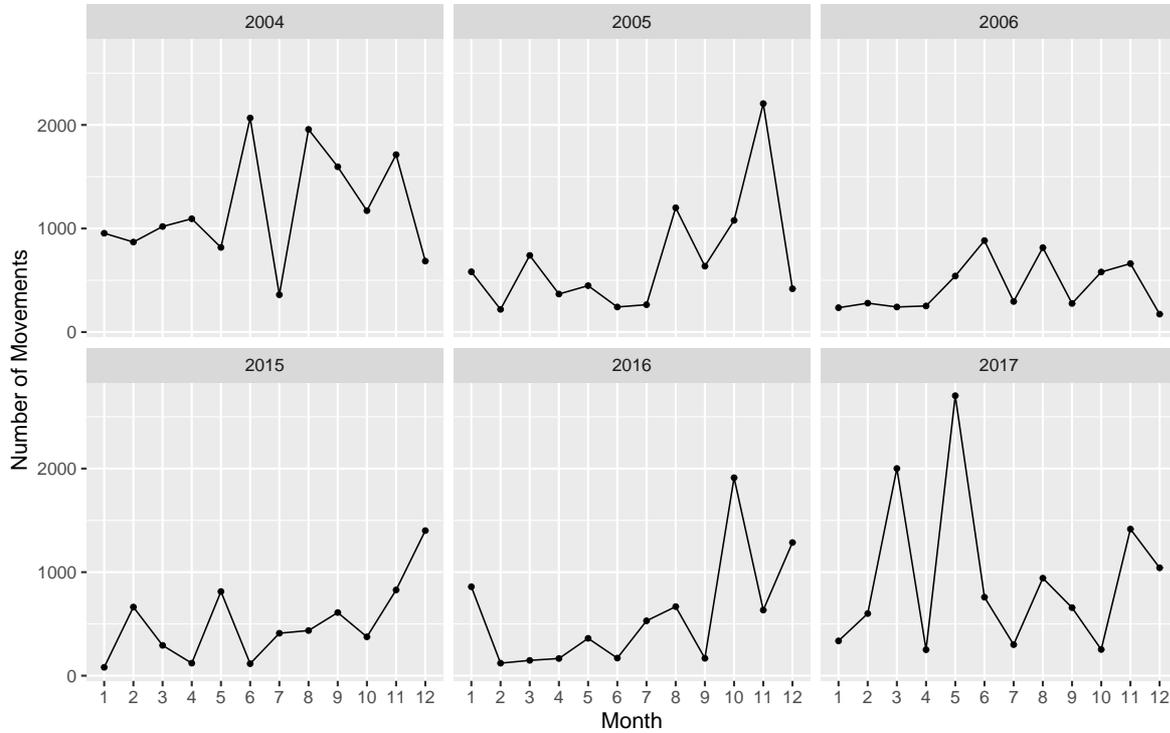}
 \caption{Number of movements per month recorded as leaving and returning to the same holding but without a short stay location recorded.}
 \label{fig:same_farm_movement}
\end{figure}

\noindent
In first exploring the CTS data to create the movement networks, we had planned to remove all Slaughterhouses, as epidemiologically they should be disease cul-de-sacs. Figure \ref{fig:num_off_slaughter} shows the number of movements we found, in the entire database, leaving nodes classified as Slaughterhouses. Having found movements \textit{leaving} Slaughterhouses we decided to keep all nodes that had out movements in our networks, regardless of type. It also led to an investigation of other possible problems with the database and whether there were differences between the two CTS updates we had access to.
\noindent
Having found movements leaving Slaughterhouses, we investigated the types of nodes the animals leaving Slaughterhouses were being taken on to. As shown in figure \ref{fig:num_off_slaughter} the majority of these were other Slaughterhouses but a number of Agricultural Holdings were also present. Figure \ref{fig:off_slaughter_overtime} shows the number of these movements off Slaughterhouses against date of movement and it clearly shows that this is becoming less and less of a problem.
\noindent
Finally, we found that some movements were recorded as leaving and returning to the same farm but not recorded with, for instance, a Market or Showground as a short stay location. This is a small number of movements as shown in figure \ref{fig:same_farm_movement} and could represent a problem in data recording. The movements in 2017 are clearly higher than the other years for which we have no explanation.
\subsection*{Differences between CTS updates}
\noindent
Having identified some problems with the data contained within the CTS database that could, at least partially, be fixed by cleaning the data, we examined two updates of the data that we had access to. Figure S11 in the supplementary information shows the differences in the number of movements off Slaughterhouses and in the number of movements which leave and return to the same holding with no short stay location - compared between the two updates of the CTS database the authors had access to. In general the more recent CTS update has slightly fewer of these types of movement (average relative percentage difference of -0.0533\% and -0.306\% respectively, the median values are 0). One simple explanation for this would be changes to how the holdings are classified. Table \ref{tab:location_changes} shows the number of locations, for the 2004 -- 2006 period, which were present in both versions of the CTS update, that had some of the ways they can be identified/grouped changed.

\begin{table}[ht]
\centering
\begin{tabular}{lr}
  \hline
  Identifier & Number of changes \\
  \hline
  CPH &   5 \\
  Location Description &   0 \\
  Holding Description & 10335 \\
  Premise Description & 206 \\
  \hline
\end{tabular}
\caption{Table number of changes to different identifiers of locations
             between updates of the CTS database for the same period (2004--2006).}
\label{tab:location_changes}
\end{table}

\noindent
\nlni
The holding description, which has changed the most between updates, was not used in the analysis of any of the data from either update of the CTS database and so we presume that alteration of location information is not the reason for the drop in ``problem'' movements.

\section*{Discussion}
\noindent
For models of disease spread or other sources of evidence to be useful for animal health policy, such evidence must be valid and credible, and the data underpinning it should be suitable, up-to-date, and well described \citep{Reeves2011}.  We have provided evidence in this report that animal movement networks, which are an important element underlying many disease models and other forms of policy support, can and do change significantly over relatively short periods of time. These changes may in--turn affect the validity or credibility of policies based on such information and preferably should be accounted for.
\nlni
Stochastic disease models based on movement networks provide estimates of risk/uncertainty that are an inherent consequence of the stochasticity \citep{Green2006}. This study demonstrates that there is \textit{structural} uncertainty due to the clear evolution of the the network metrics in our network over this time period. Therefore any risk--based evidence from network modelling should consider presenting risk arising from both the stochasticity and the uncertainty between the recorded network used for modelling and the real network at the time of the outbreak. Presenting such risks may also entail presenting a larger range of probable outcomes to decision-makers, whilst reducing the uncertainty in the network used may lead to the opposite. Such considerations are important because network analysis and epidemiological models can provide important analysis for industry if presented appropriately \citep{Dube2009}.
\nlni
For researchers interested in networks or modelling disease spread through the UK cattle system \citep{Woolhouse2005, Green2006, Kao2007, Green2008, Vernon2009} the CTS database, as it contains details on so many cattle movements, is an excellent resource. \citet{ShirleyRushton2005EpiInf}, for example, investigated the network of infected holdings concluding that the strategy used to control the FMD outbreak in 2001 was not as effective as more targeted measures could have been, but was all that was possible after the disease was discovered. The CTS dataset allows clear identification of markets and enables researchers to either remove them from their networks or include them. \citet{Robinson2007b} examined cattle movements from 2002 -- 2004 to investigate the role of markets in disease spread and conclude that, as it is also dependent on how quickly the disease may be transmitted, it ``needs consideration on a disease by disease basis''.
\nlni
When comparing the monthly networks constructed from CTS data across our two time periods, it is clear that there have been substantial changes as demonstrated both graphically and quantitatively by formal statistical modelling. In our network measures of total nodes with movements, total movements, births and deaths, we have shown reductions in the more recent networks. The decreases in the size of these measures seems to mirror the trends in the figures published by government \citep{scot_gov_industry, uk_gov_industry} which show a decline in the number of holdings. The seasonal peaks of births within the year are also similar to those found previously and represent individual peaks for the beef and dairy herds \citep{Gates2013b}. The causes of these changes (and those of the network metrics) could come from an extremely wide range of factors including wider changes in farmer behaviour, economic considerations, trade and import restrictions or disease prevention \citep{Vernon2012, Gates2015} but are outwith the scope of this work.
\nlni
%Metrics
The results show statistically significant differences in six network metrics. The reductions in the proportion of nodes in GSCC suggests a smaller lower bound on the maximum epidemic sizes in 2015 -- 2017 compared to 2004 -- 2006 \citep{Kao2009}. The result for the GWCC suggests the same for the upper bound. The slight reduction in assortativity, although not significant, means the network has become slightly more dissassortative, so holdings with high degree values tend to send/receive animals from holdings with lower degree values. Disease spreads slower on dissassortative networks \citep{Kao2007} and their GSCC can be reduced rapidly by removing nodes of high degree \citep{Newman2003a, Natale2009, Mweu2013}. This suggests that an epidemic on the networks in 2015 -- 2017 might spread more slowly than through the earlier network and it could be easier to reduce the, already smaller, lower bound on maximum epidemic size. The decrease in the median value of betweenness between time periods is a possible indication that the network is a little more spread out. If there are more bridges, or bridges are circumvented (a connection between nodes $E$ and $C$ in figure \ref{fig:betweenness_network} for example) then the overall median value of betweenness decreases. The decrease in the median total degree provides evidence that holdings are trading animals with a smaller number of other holdings between 2015 -- 2017 compared to 2004 -- 2006.
\nlni
As we had access to two updates of the CTS data, we compared the frequency of some ``problem'' movements from 2004 -- 2006 between these updates. The number of each ``problem'' movement decreased, indicating that historical data is amended in the more recent update of the CTS database. The numbers of movements leaving Slaughterhouses are very small (average of 0.0487\% across 2004 -- 2006 and 2015 -- 2017) when compared to the total number of animal movements in the monthly networks, so it may just be a data cleanliness issue. However, there are several other possible reasons for this including health issues discovered upon arrival or if there were difficulties with documentation \citep{Foddai2019}. \citet{Knific2020} has shown that using more recent data gave better results when trying to reduce the GSCC via targetting node removal by values of betweenness or total degree. These issues need to be taken into consideration by anyone planning on using the CTS database in a future study. The authors suggest that similar considerations should be applied to similar datasets from other countries e.g. Canada \citep{Dube2008}, Germany \citep{Buttner2013, Buttner2014}, Italy \citep{Natale2009, Bajardi2011}, Sweden \citep{Lindstrom2009, Frossling2012} and Switzerland \citep{Vidondo2018}.
\nlni
Of two updates of the CTS data, the first contained all movements from 1$^\textrm{st}$ January 2001 until 14$^\textrm{th}$ February 2014, whilst the second, contained all movements from 1$^\textrm{st}$ January 2001 up to 31$^\textrm{st}$ December 2017. Instructions for entering data into the CTS are available online \citep{bcms-cts} and include the information that all movements should be reported to the British Cattle Management Scheme (BCMS) within three days of the movement. In future, expediting the process of making large datasets (or just updates) available to researchers would be beneficial when differences due to updating, like those shown above, may be present.
\nlni
When constructing our networks, all nodes with animals leaving them were included, for the reasons outlined in section 2.1. This meant that some Slaughterhouses and all Markets were included, both of which (amongst other holding types) have been removed from previous studies \citep{Woolhouse2005, Vernon2009, VanderWaal2016}. Similar consideration should also be made to the length of stay. For epidemiological reasons, earlier studies have ignored movements where the animal has a short stay \citep{Woolhouse2005, Vernon2009} - normally at a Market or Showground type location - as stays of this length may not present sufficiently long exposure time to contract certain diseases. In comparison Markets and Dealers have been identified as important \citep{Ortiz2006, Robinson2007b, Rautureau2011} in epidemics, including FMD. \citet{Kao2007} have already examined this idea of endemic disease versus an epidemic and the inclusion/exclusion of movements involving markets.
\nlni
Similarly, when modelling disease, there should be a consideration of whether to use dynamic or static networks. The CTS database has already been used to examine these different approaches \citep{Vernon2009}. A dynamic network was compared to various 7 or 28 day static networks. When comparing the results of the static networks, \citep{Vernon2009} found little difference between 7 and 28 days except for a smaller final epidemic size on the 7 day networks. Dynamic networks are, ideally, the preferred method of modelling disease spread, be it within herd or between herd disease spread \citep{Vernon2009, Duncan2012} but considerations also need to be made for which type of disease spread is being modelled (e.g. endemic or an epidemic) \citep{Kao2007}.
\nlni
Within this work the focus was on the properties of the network itself, rather than modelling disease spread, and static networks were used to approximate the dynamic cattle movements. These properties are described by network metrics, some of which only make sense when considering a static network over a period of time. However, as we have shown, the CTS network has changed in some way between 2004 -- 2006 and 2015 -- 2017 and it has been over ten years since the last FMD outbreak. This suggests that not only is it worth considering revisiting such modelling work but also raises the question of how often models should be recalculated with the most recent available network data. The impact of the changes in network measures and metrics on epidemic size will be unknown, until this modelling is conducted. This extends to whether the results from the statistical models are effective predictors of the importance of the measures and metrics. That is, do significant differences in network metrics imply significant differences in disease spread. Depending on the disease modelled, the differences observed in some network metric (e.g. GSCC, GWCC, betweenness and strength) may be more or less important than the small changes observed in others (e.g. assortativity, average path length and reciprocity). In some cases this modelling is happening anyway as part of live exercises \citep{epic2019} but it deserves wider consideration. Tools \citep{Boettiger2015, Kurtzer2017} also exist to allow for reproducible pipelines to be created in advance of a disease outbreak. These could be utilised to provide summary statistics for policy makers or to enable modellers to rerun their calculations on current data, as and when they were required. Partial data has been used to predict disease spread with (for example) swine movements \citep{Valdes2017} and networks already reconstructed from partial CTS data \citep{Dawson2015} to consider FMD. However, what the CTS data should offer is a complete historical record of cattle movements where network predictions can be based on many years of movement data. Such an approach may allow for an extrapolation of the trends we see in networks so as to allow for more accurate disease modelling prior to receipt of complete movement data.

\subsection*{Conclusions}
\noindent In this paper we have shown that monthly networks of cattle movements constructed using data from the Cattle Tracing System have altered in structure between the period 2004 -- 2006 and the most recent calendar years available for analysis, the period 2015 -- 2017. They have changed both in terms of simple network measures (total number of nodes with movements, number of movements, births and deaths) and in terms of network metrics. The decreases shown in GSCC and GWCC indicate that, all else equal, epidemic sizes would be smaller in 2015 -- 2017, compared to 2004 -- 2006. By comparing different updates of the CTS database covering the same time period we were also able to show evidence of historical data being amended in the intervening period. Accurate modelling of disease spread through a network requires representative descriptions of the network. For both the above reasons (evolution of the network and amending of data) it is clear that older data will not be as representative as recent data. Therefore the authors recommend that the most recent available data always be used for network construction and analysis. A consequence of this is that the data must be made available to stakeholders as soon as possible to ensure that analysts can provide relevant interpretation to inform decision--making in a timely and robust manner. The Cattle Tracing System provides an invaluable resource of historical cattle movement data that could be utilised to predict cattle movement network metrics and possibly movement data for time periods where data is not yet available.

%\section*{List of Abbreviations}
%
%\begin{itemize}
%\item Cattle Tracing System (CTS)
%\item Foot and Mouth Disease (FMD)
%\item United Kingdom (UK)
%\item British Cattle Management Scheme (BCMS)
%\item county parish holding (CPH)
%\item UK Animal and Plant Health Agency (APHA)
%\end{itemize}

\section*{Author Contribution}

\noindent The contribution of the authors is as follows. Andrew J Duncan: Conceptualisation, Formal Analysis, Visualisation, Writing – Original Draft/Review \& Editing. Aaron Reeves: Data Curation, Supervision, Writing - Review \& Editing.
George J Gunn: Writing - Review \& Editing. Roger W Humphry: Conceptualisation, Supervision, Writing - Original Draft/Review \& Editing.

\section*{Conflict of Interest Statement}
%All financial, commercial or other relationships that might be perceived by the academic community as representing a potential conflict of interest must be disclosed. If no such relationship exists, authors will be asked to confirm the following statement:

\noindent The authors declare that the research was conducted in the absence of any commercial or financial relationships that could be construed as a potential conflict of interest.

\section*{Funding}

\noindent The study  was funded by the Scottish Government’s Rural Affairs, Food and the Environment Strategic Research Portfolio 2016-2021, as part of the Strategic Research Programme 2016-2021 and the Centre of Expertise on Animal Disease Outbreaks (EPIC). SRUC is one of the Scottish Environment, Food and Agricultural Institutes (SEFARI).

\section*{Acknowledgments}
\noindent The authors gratefully acknowledge the assistance of the Animal and Plant Health Agency, in particular Jon Weston, Gareth Hateley and Alessandro Foddai for their invaluable assistance, comments and provision of CTS data.

\section*{Supplementary Material}
\noindent All supplementary material is contained in \textit{SupplementaryMaterial.pdf}.

\section*{Data Availability Statement}
\noindent The data that support the findings of this study are available from APHA but restrictions apply to the availability of these data, which were used under license for the current study, and so are not publicly available. Data are however available from the authors upon reasonable request and with permission of APHA.

\bibliography{cts_final.bib}

\end{document}